\documentclass{jfm}

\usepackage{bm,amsmath,hyperref,cleveref}
\usepackage[OliveGreen,dvipsnames]{xcolor}
\usepackage[normalem]{ulem}

\def\NEW#1{\textcolor{red}{#1}}
\def\NEW#1{\textcolor{OliveGreen}{#1}}
\def\NEW#1{\textcolor{cyan}{#1}}
\def\NEW#1{\textcolor{black}{#1}}

\shorttitle{DSS using GCE2}
\shortauthor{G. V. Nivarti, J. B. Marston and S. M. Tobias}

\title{Direct Statistical Simulation using generalised cumulant expansions}

\author{G. V. Nivarti\aff{1}\corresp{gvn22@cantab.ac.uk} \and J. B. Marston\aff{2} \and S. M. Tobias\aff{1}}
\affiliation{\aff{1}Department of Applied Mathematics, University of Leeds, Leeds, UK \aff{2}Brown Theoretical Physics Center and Department of Physics, Brown University, Providence, RI 02912-S USA}

\begin{document}

\maketitle

\begin{abstract}
In recent years, the Generalised Quasilinear (GQL) approximation has been developed and its efficacy tested against purely quasilinear (QL) approximations. GQL systematically interpolates between QL and fully non-linear dynamics by employing a generalised Reynolds decomposition. Here, we examine an exact statistical closure for the GQL equations on the doubly periodic $\beta$-plane. Closure is achieved at second order using a generalised cumulant approach which we term GCE2. GCE2 is shown to yield improved performance over statistical representations of purely QL dynamics (CE2) and thus enables Direct Statistical Simulation (DSS) of complex mean flows that do not entirely fall within the remit of pure QL theory.  \NEW{Despite the existence of an exact closure, GCE2 like CE2 admits the  possibility of a rank instability that leads to differences with statistics obtained from GQL.  Recognition of this instability is a necessary step before further progress can be made with the GCE2 statistical closure.}
\end{abstract}
 
\begin{keywords}
DSS; QL; CE2; GQL\NEW{; GCE2}.
\end{keywords}

\section{Introduction}

Fluid turbulence, where nonlinear interactions occur over a wide range of spatial and temporal scales, plays an important role in engineering, geophysical, astrophysical and even biological fluid mechanics. Much turbulence research focuses on the idealised case of homogeneous and isotropic turbulence, despite the canonical situation involving important inhomogeneities and anisotropies. For example, in geophysical and astrophysical situations rotation and stratification may play an important role in selecting preferred directions, whereas in other cases mean flows and boundaries often lead to both inhomogeneities and anisotropies. For these cases, it is important to develop a framework that builds in inhomogeneity and anisotropy from the outset and turns this ``bug" into a ``feature''. Such a framework involves constructing equations for the evolution of the statistics of the turbulence; it is important to bear in mind that the presence of anisotropy and inhomogeneity often leads to non-trivial low-order statistics --- for example, sustained mean flows that interact strongly with fluctuations. Methods designed for describing the evolution of such flows \textit{will} perform badly for the homogeneous isotropic case, where mean flows are absent. For a description of the many methods that have been developed and the underlying philosophy of this approach see the review by \citet{mt_2023}.

In general these methods rely on developing equations governing the evolution of the low-order statistics for the flow (often termed the cumulants). Such an approach often requires a closure approximation, where the higher order statistics are either neglected completely or written as functions of the low-order cumulants. However, if the system exhibits quasilinear dynamics, then the system of low order statistical equations closes exactly \citep{Herring:1963} and no further approximations are needed; such a system of statistical equations is known as CE2 (representing a cumulant expansion at second order). Recent years has seen many systematic investigations of the validity of the quasilinear approximation (QL) in representing the full nonlinear dynamics \citep[see e.g.][]{Tobias2013,chmt2016,Marston2019}; it has been determined that there are certain circumstances where the QL approximation breaks down and better approximations are needed. 
 
\NEW{Recent research has focused on a generalisation of the quasilinear approximation (termed GQL) that generalizes the definition of the mean flow to include long-wavelength modes in the zonal direction in addition to the spatial mean \citep{mct2016}. A scale separation into long wavelength modes that interact fully nonlinearly with themselves, and short wavelength modes with wavenumbers greater than a cutoff $\Lambda$ that interact only quasilinearly with the long wavelength modes, has certain advantages in that it allows energy to be scattered among the turbulent eddies through interaction with the generalized mean flow. This eddy-scattering often leads to a more faithful representation of the nonlinear dynamics than QL, where these interactions are forbidden.} 
 
\NEW{GQL, like QL, is a conservative approximation but one that systematically interpolates between QL (when $\Lambda = 0$) and NL. As $\Lambda \rightarrow \infty$ the GQL system consists solely of fully interacting low modes and reverts to NL, albeit not necessarily monotonically. In particular, if the cutoff $\Lambda$ is large (but not infinite) it is possible that high modes will be stable and (incorrectly) not have any energy \citep{hern_2022a}.}

In this paper we derive the statistical representation of the GQL approximation (which we term GCE2) and describe the utility of this approach by comparing it with CE2 for two model fluid dynamics problems (one deterministic and one stochastic) describing the interaction of mean flows with turbulence in two dimensions.

\NEW{Recently it has been shown that a rank instability can occur in CE2 leading to a divergence in statistics from QL (\cite{Nivarti2022}; see also \cite{Oishi2022}).  The existance of this instability has apparently been missed in all prior work going back to \citep{Herring:1963}, and the resolution presented here now allows us to properly compare GCE2 to statistics obtained from GQL and compare both to the statistics found in QL and fully nonlinear (NL) simulations.}

\section{The Generalised Quaslinear Approximation and Its Closure}

\NEW{The Generalized Quasilinear (GQL) approximation has been studied in a variety of fluid contexts by a number of different groups.  We refer the reader to our review article \citep{mt_2023} for an introduction to the approximation and its physical interpretation.}

\subsection{The Generalised Quasilinear (GQL) Approximation}
We consider a system of non-linear dynamical equations for a state vector $\bm{q}(\bm{x},t)$  written as
\begin{equation}
    \partial_t \bm{q} = \mathcal{L}[\bm{q}] + \mathcal{N}[\bm{q},\bm{q}] ,
    \label{eq:nl}
\end{equation}
with $\mathcal{L} [\cdot]$ a linear and $\mathcal{N}[\cdot,\cdot]$ the non-linear (in this case, quadratic) operator. In order to apply the GQL approximation, the state vector $\bm{q}$ is expanded using a spectral basis along the zonal direction (more generally, the direction exhibiting statistical homogeneity). GQL then proceeds \citep{mct2016} by applying a low-pass filter (projection operator) with cutoff $\Lambda$ in the zonal direction, leading to a generalised Reynolds decomposition of the state vector
\begin{equation}
    \bm{q} = \bm{q}_\ell  + \bm{q}_h ,
    \label{eq:gr}
\end{equation}
where the subscripts ($\ell$) and ($h$) denote low and high zonal wavenumber modes, respectively. 
For the two dimensional Cartesian models considered here

\begin{equation}
\bm{q}_\ell 
=\sum_{k=-\Lambda}^{\Lambda}{\bm{q}}_{k}(y) \,e^{i k x} \quad \quad
\bm{q}_h = \bm{q} - \bm{q}_\ell.
\label{GQL}
\end{equation}

This decomposition obeys the usual rules of orthogonality and idempotence, and can be simplified to the conventional Reynolds decomposition (into mean and fluctuation) simply by setting $\Lambda = 0$. {Note, however, that the conventional Reynolds decomposition obeys the equality $\overline{\overline{ \bm{q} }  \bm{q}} = \overline{\overline{ \bm{q} } \, \overline{\bm{q}}}$. {Note also that we use the overline to indicate a mean computed as a spatial (zonal) average, but other averages are also possible including ensemble and time averages.)}. In comparison, $({\bm{q}_\ell \bm{q}})_\ell \ne (\bm{q}_\ell\bm{q}_\ell)_\ell$ for $\Lambda > 0$ under the generalised Reynolds decomposition into low- and high-modes; this inequality is remedied after applying certain interaction rules as follows.}

Applying the decomposition \ref{eq:gr} to the $\bm{q}$ in \cref{eq:nl}, gives rise to various classes of non-linear terms that correspond to different triadic interactions involving low- and high- modes. The possible triad interactions can be represented using triadic diagrams in which low-modes (zonal wavenumber $|m| \leq \Lambda$) and high-modes $|m| > \Lambda$ are denoted using low-frequency and high-frequency wave edges \citep{mct2016,mt_2023}. GQL equations retain non-linear self-interactions giving rise to low modes, as well as quasilinear interactions between low- and high-modes. {Non-linear self-interactions giving rise to high-modes and interactions between low and high modes giving rise to low modes are both dropped \cite[see the diagrams in Sec 2 of ][]{tom_2018}}. \NEW{This elimination of triad interactions, as in the case of triad decimation by pairs \citep{Kraichnan1985}, conserves quadratic invariants such as energy and enstrophy.} The following GQL equations are obtained by applying these \emph{interaction rules} 
\begin{align}
    \label{eq:gql:low}
    \partial_t \bm{q}_\ell & = \mathcal{L}[\bm{q}_\ell] + \mathcal{N}_\ell[\bm{q}_\ell,\bm{q}_\ell] + \mathcal{N}_\ell[\bm{q}_h,\bm{q}_h], \\
    \label{eq:gql:high}
    \partial_t \bm{q}_h & = \mathcal{L}[\bm{q}_h] + \mathcal{N}_h[\bm{q}_h,\bm{q}_\ell] + \mathcal{N}_h[\bm{q}_\ell,\bm{q}_h] ,
\end{align}
for the low- and high-modes. Here, $\mathcal{N}_\ell[\cdot, \cdot]$ is the projection of the nonlinearity onto the low modes and $\mathcal{N}_h = \mathcal{N}[\cdot, \cdot] - \mathcal{N}_\ell[\cdot, \cdot]$, which denotes the high-mode spectral projection of the non-linear operator $\mathcal{N}[\cdot, \cdot]$.  Furthermore,  when $\Lambda = 0$, we have $\bm{q}_\ell = \overline{\bm{q}}$ and $\bm{q}_h = \bm{q}^\prime$, and the GQL equations reduce to the well-known system of QL equations \citep{Marston2019}. Similarly, setting $\Lambda$ as the full spectral resolution in the zonal direction, we obtain $\bm{q}_\ell = \bm{q}$ and $\bm{q}_h = 0$; the GQL equations are identical to the fully-nonlinear (NL) equations \cref{eq:nl} in this limit. Hence, the GQL equations interpolate systematically {(though not monotonically)} between QL and NL dynamics by varying the zonal spectral cutoff $\Lambda$. Crucially, the GQL equations lack high-mode non-linearities which makes GQL amenable to statistical closure as \cref{eq:gql:high} is formally linear in $\bm{q}_h$. We shall make use of this property now.
%
%

\subsection{Deriving the Generalised Cumulant Expansion at Second Order (GCE2)}

Statistically-closed equations termed CE2 have been derived for quasilinear (QL) equations using cumulant expansions and other methods \citep[see e.g.][]{Farrell2007,marston65conover,Farrell2013,Constantinou2015}. A closure for QL equations is achieved at second order -- the equations for the mean and fluctuation terms are used as a starting point for deriving the corresponding equations for the first two cumulants: $\overline{\bm{q}}$ and $\overline{\bm{q}^\prime\bm{q}^\prime}$, respectively. This strategy allows for Direct Statistical Simulation (DSS) of low-order statistical quantities that correspond to QL dynamics. In a similar manner, statistical closure for GQL dynamics is also achievable at second order using generalised cumulant expansions which employ the notion of the mean implicit within the generalised Reynolds decomposition of \cref{eq:gr}. We thus define the first two generalised cumulants as 
\begin{align}
    c_1 & \equiv \bm{q}_\ell, \nonumber \\
    c_2 & \equiv \bm{q}_h \bm{q}_h, 
\end{align}
following the spectral filter (projection operator) notation used earlier. Closed form equations for these generalised cumulants, termed GCE2, are obtained by following a similar approach to that adopted for deriving CE2 equations from QL equations \citep{Marston2019}. The generalised first cumulant is identical to low modes (analogous to $\overline{\bm{q}}$ in CE2). As a result, the equation governing the evolution of the generalised first cumulant $c_1$ is the same as that for the low modes $\bm{q}_\ell$ as given by \cref{eq:gql:low}: 
\begin{align}
    \label{eq:gce2:c1}
    \partial_t c_1 & = \mathcal{L}[c_1] + \mathcal{N}_\ell[c_1, c_1] + \mathcal{N}_\ell[c_2].
\end{align}
The generalised second cumulant is a field bilinear in high modes (akin to the Reynolds stress $C = \overline{\bm{q}^\prime\bm{q}^\prime}$ in CE2). The equation for $c_2 = \bm{q}_h \bm{q}_h$ is thus obtained by multiplying the high-mode equations \cref{eq:gql:high} with high-modes and subsequently projecting down to the low-modes as follows
\begin{align}
    \partial_t (\bm{q}_h \bm{q}_h) & = \{\bm{q}_h \partial_t \bm{q}_h\}  = \{\bm{q}_h \mathcal{L}[\bm{q}_h]\} + \{\bm{q}_h \mathcal{N}_h[\bm{q}_h,\bm{q}_\ell]\},
    \nonumber\\
    \implies \partial_t c_2 & = \mathcal{L}[c_2] + \{\mathcal{N}_h[c_2, c_1]\}. 
    \label{eq:gce2:c2}
\end{align}
where the curly braces $\{\cdot  \cdot \}$ denote symmetrization, i.e. $\{a b\} = a b + b a$. Taken together, \cref{eq:gce2:c2} and \cref{eq:gql:low} form a closed set of equations in terms of generalised cumulants. This allows the implementation of Direct Statistical Simulation of GQL dynamics, i.e. solve directly for statistics that interpolate between QL and NL dynamics. \NEW{We note here that similar equation sets have been derived in other contexts  \citep[see e.g.][]{bakasioannou2013b, bakasioannou2013c, Constantinou2015, Constantinou:2016fp}, though the relationship with generalised quasilinear models and the possibility of the statistical models diverging from the QL-type models has not been investigated previously.}

A recent study has demonstrated that although one can mathematically arrive at the CE2 system starting from the QL equations, the two frameworks are not statistically equivalent \citep{Nivarti2022}. This is because, as opposed to QL equations, the CE2 system can sustain solutions wherein the second cumulant has rank $> 1$ whereas the second cumulant obtained by zonally averaging QL dynamics is always of unit rank. We anticipate therefore that the GCE2 and GQL systems may not be fully equivalent either. However, we emphasise that the ability of GCE2 to predict a generalised second cumulant with rank greater than unity ought to be considered as a feature not a bug, one that motivates further study.

This paper has two aims: 1) to test the predictions of GCE2 against those of GQL, thereby demonstrating that, even for a cutoff $\Lambda = 1$, GCE2 improves significantly over CE2, and 2) to probe observed divergences between GCE2 and GQL, relating them to the presence of rank instabilities as in the case of CE2 and QL \citep{Nivarti2022}.  \NEW{We employ only low resolution models here with a small number of modes to highlight clearly any differences between GQL and GCE2.  We have verified that the qualitative results continue to hold at higher resolutions as well.}

\section{Numerical Implementation}
We conduct simulations of a rotating, incompressible fluid on a doubly-periodic $\beta$-plane. The time evolution of the relative vorticity $\zeta \equiv \widehat{z}\cdot (\nabla \times {\boldsymbol u})$ is given by 
\begin{equation}
    \label{eq:vorticity}
    \partial_t \zeta = \beta \partial_x \psi - \kappa\zeta + \nu \nabla^2 \zeta + J[\psi,\zeta] + F, 
\end{equation}
where ${\boldsymbol u}$ is the velocity, $J[\psi,\zeta] = \partial_x\psi \partial_y \zeta - \partial_x \zeta \partial_y \psi$ and the streamfunction $\psi \equiv \nabla^{-2}\zeta$. Gradients of rotation are included via the $\beta$ term, `bottom friction' via the $\kappa$ term and viscosity via the $\nu$ term. The forcing term $F$ models energy injection into the system. We adopt two different models of forcing (or driving) in this study: 1) a deterministic steady two-scale Kolmogorov-type forcing as used by \cite{Tobias2017} - see equation~(\ref{eq:kf}), and 2) a white-in-time stochastic model of forcing adapted from \cite{Constantinou2016} - see equation~(\ref{eq:sf}). The latter is a commonly-used model that mimics thermal driving.

The spectral solver \verb+ZonalFlow.jl+ (written in Julia \citep{Julia2017} and made available online \citep{Nivarti2021}) is used to obtain direct numerical solutions of equation (\ref{eq:vorticity}). Timestepping algorithms are imported from the well-tested ecosystem of the \verb+DifferentialEquations.jl+ \citep{Rackauckas2017} package. We use different timestepping methods to simulate flows with different types of driving models. For the deterministically-driven Kolmogorov flow, we use the explicit $5/4$ Runge-Kutta method of Dormand-Prince with a fixed timestep of $\Delta t = 0.001$. For the stochastically-driven flow, we use the SRIW1 method of order $1.5$. To be consistent, we use the same timestepping algorithm to solve the dynamical equations (QL/GQL/NL) and the statistical equations (CE2/GCE2) for a given flow. Unless specified otherwise, the initial condition consists of random noise of mean power $10^{-4}$. We employ a $[0,2\pi]^2$ grid with resolution \NEW{$M \times N = 10 \times 10$ for the Kolmogorov flow system, and a $[0,2\pi]\times [0,\pi]$ grid with a resolution of $12 \times 20$ for the stochastically-driven system, where $|m| < M$ is the wavenumber in the zonal ($x$) direction and $|n| < N$ is the wavenumber in the non-zonal ($y$) direction.}  

We have conducted a number of validation tests to ensure that the equations are implemented accurately in our code. {Solutions for equations neglecting the dissipative terms have been tested to confirm conservation of energy and enstrophy is satisfied in such regimes.} Additionally, the agreement of mean equilibrium solutions for a given energy input with dissipation has also been confirmed across the various equation systems. In the presence of non-linear terms, our tests confirm that setting the spectral cutoff $\Lambda = 0$ in GCE2 reproduces CE2 results and setting \NEW{$\Lambda = M - 1$} reproduces the fully-nonlinear (NL) results. In addition to the forcing models outlined above, the aforementioned tests are also conducted for a determinstic driving model with relaxation to a point jet \citep{Marston2008}; here, an additional validation test is available as the GCE2/CE2 as well as GQL/QL results converge to NL for small relaxation times. All these tests have been automated, such that the Github workflows server conducts them each time a code change is pushed to the online repository \citep{Nivarti2021}.

In the following, we present results comparing GCE2 against GQL for a spectral cutoff $\Lambda = 1$. For each forcing model and specified set of parameters, we compare these results against predictions of the NL system as well as of the CE2 and QL equations. Note that, we use overlines to indicate spatial (zonal) averages (not ensemble or time averages) on QL solutions as in the comparisons against CE2. When time averages are used (such as to facilitate comparison of H\"ovm\"oller plots), we explicitly state this along with specifying the averaging window in the accompanying text.

\section{Results}

\subsection{Two-scale Kolmogorov forcing}
\begin{figure*}
    \centering
    {\includegraphics[scale=0.6]{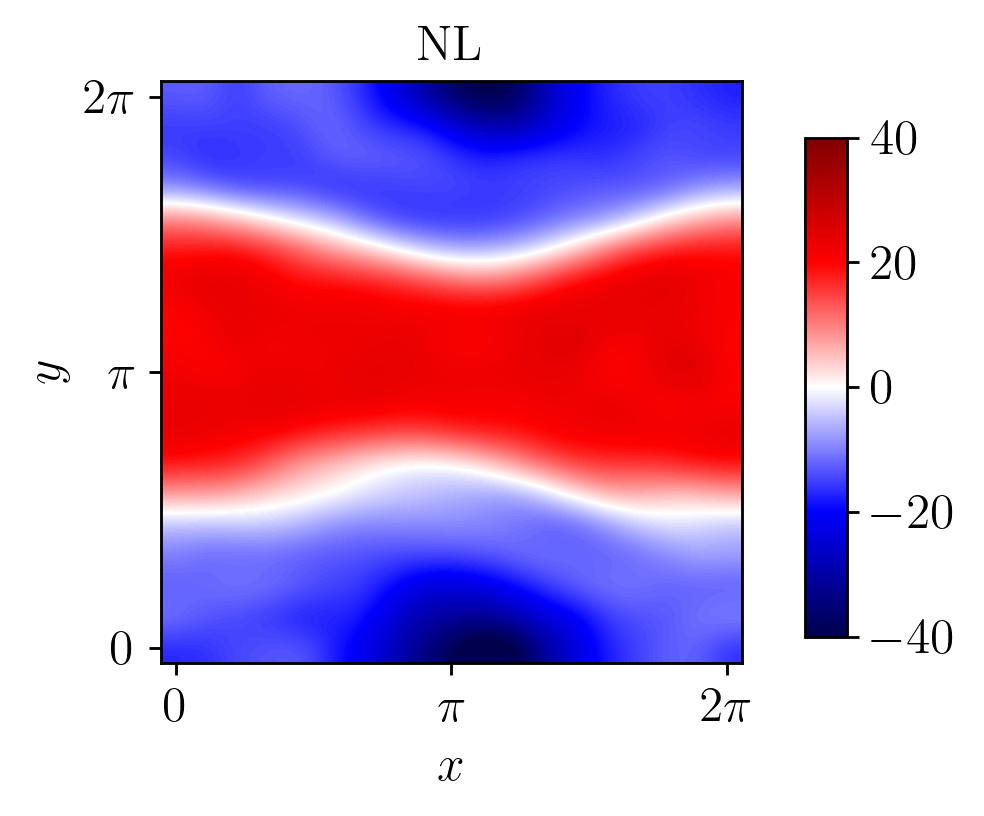}}\\
    {\includegraphics[scale=0.6]{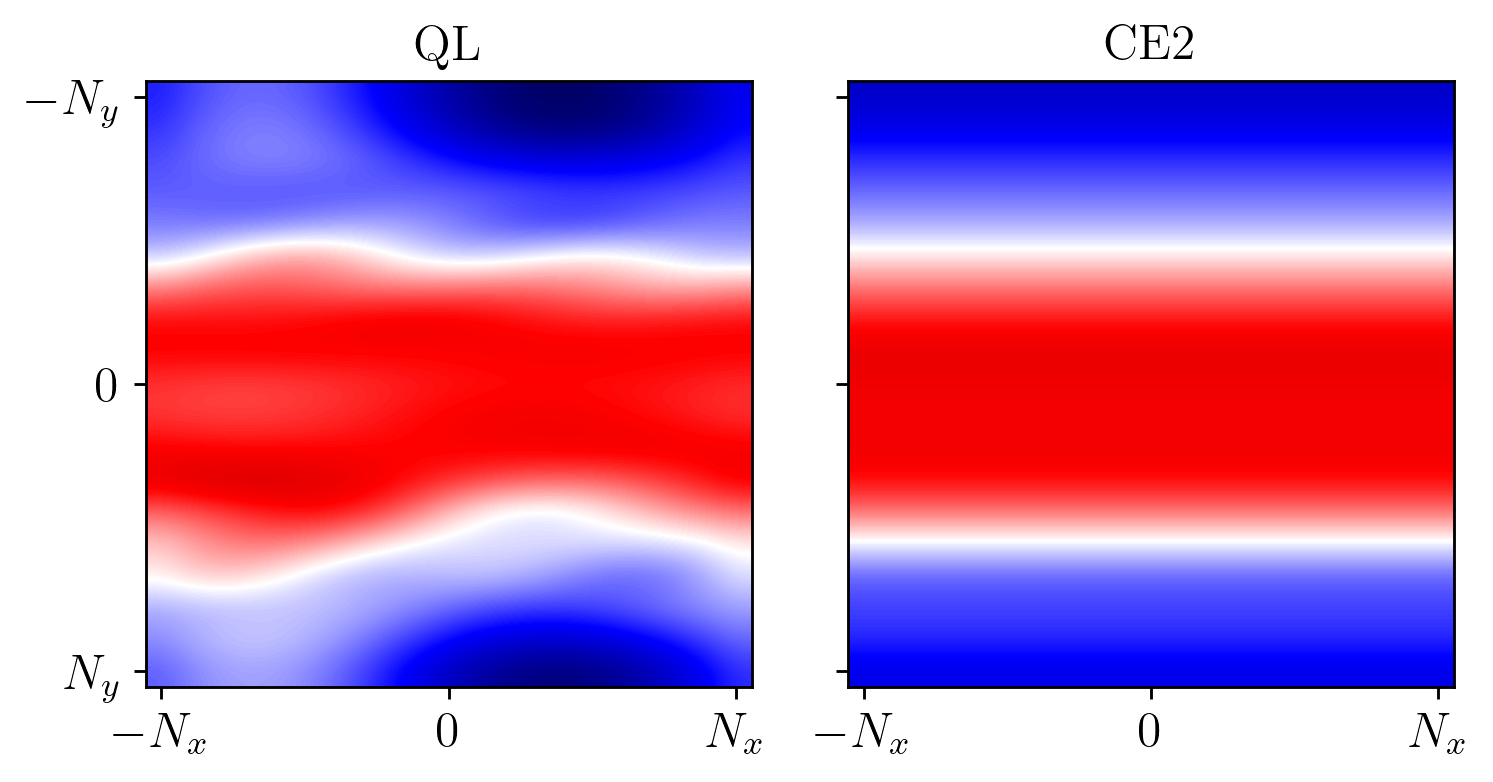}}\\
    {\includegraphics[scale=0.6]{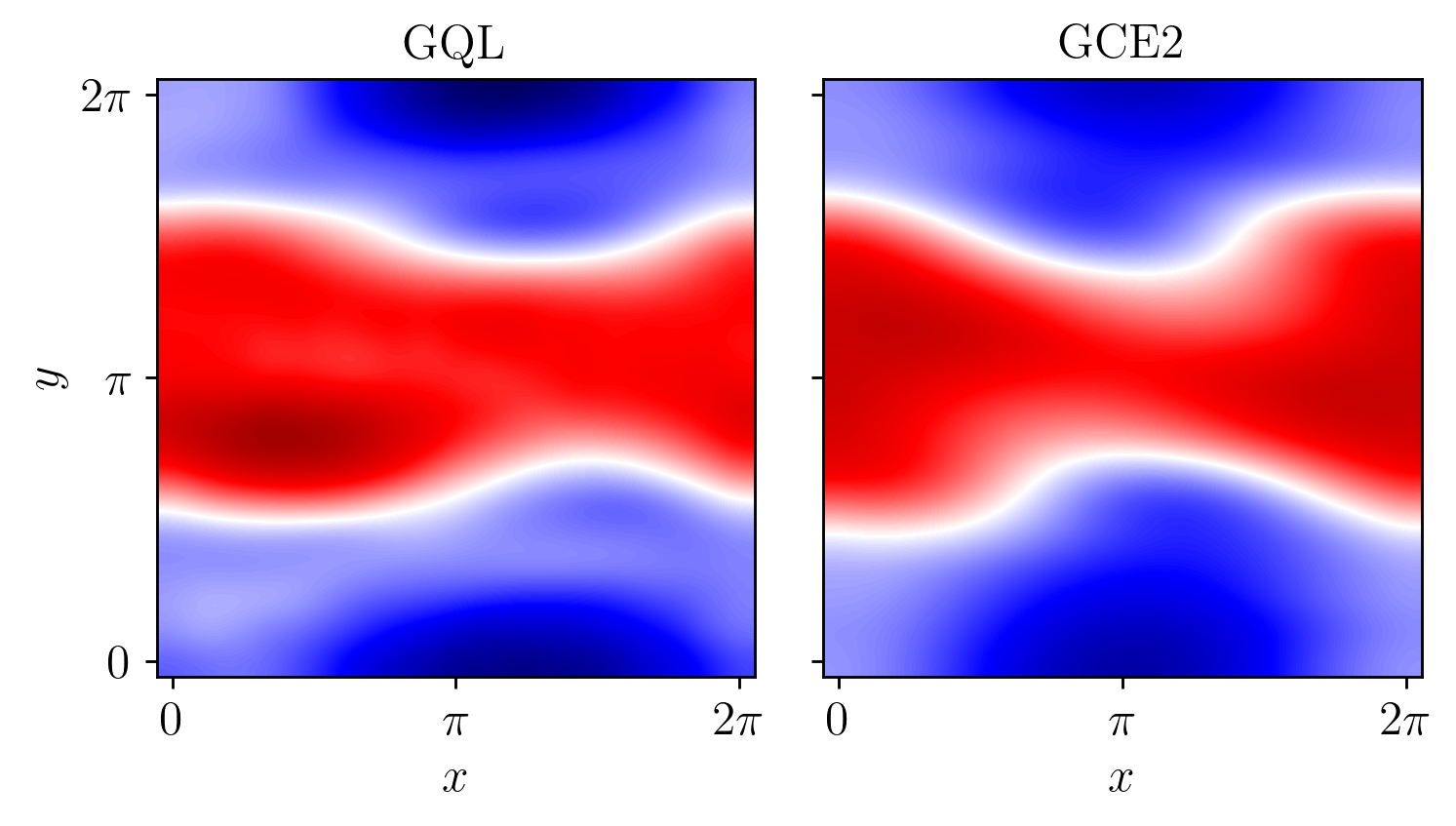}}
    \caption{Snapshot of the voriticity field  ${\zeta(x,y)}$ for NL (top), QL versus CE2 (middle) and GQL versus GCE2 (bottom) for two-scale Kolmogorov flow with resolution $M = N = 10$ at $t = 1000$ days. All colour ranges are identical.}
    \label{fig:kf:vorticity}
\end{figure*}

\Cref{fig:kf:vorticity} shows the final time ($t = 1000$ days) resolved vorticity solution for a flow driven by a deterministic two-scale Kolmogorov forcing.  We set $F$ in \cref{eq:kf} to be
\begin{equation}
F(y) = -\rm{cos}(y) -8\rm{cos}(4y).
\label{eq:kf}
\end{equation}
and the parameters as $\nu = 0.02$ and $\beta = \kappa = 0$ in \cref{eq:vorticity}. This system is known to lead to non-trivial dynamics \citep{Tobias2017}. \NEW{The forcing term is added to the tendency equation for the first cumulant, Eq. \ref{eq:gce2:c1}.}

The NL solution (top panel in \cref{fig:kf:vorticity}) consists of a strong band of positive vorticity centred at $y = \pi$ surrounded by negative vortical regions, as was observed by \cite{Tobias2017}. The band of positive vorticity appears to be composed of a strong zonal mean and a sinusoidal zonal wave with wavenumber $m = 1$, the latter contributing to the varying width of the band as a function of $x$. As noted in \cite{Tobias2017}, this vorticity solution corresponds to a rightward flow on the upper half $y> \pi$ of the domain and leftward flow in the lower half.

The QL solution (middle left panel in \cref{fig:kf:vorticity}) captures the central positive vorticity band, albeit with an apparently weaker $m = 1$ harmonic resulting in a band of more uniform height and thickness. {For presentation of results, we adopt zonal (spatial in $x$ direction) averaging so that $\overline{\bm{q}}$ denotes the zonal mean of quantity $\bm{q}$.} The zonal mean vorticity $\overline{\zeta}$ predicted by CE2 (middle right) agrees reasonably well with QL dynamics both in magnitude and distribution. The localised negative vorticity regions predicted by QL (middle left panel) in the upper and lower half of the domain are averaged out in the CE2 prediction (middle right panel). The bottom panels in \cref{fig:kf:vorticity} show the GQL (left) and the resolved GCE2 (right) solutions for a cutoff $\Lambda = 1$. These are in very close agreement with the fully-nonlinear solution, improving significantly over the QL/CE2 solutions. 

\begin{figure*}
    \centering
    {\includegraphics[scale=0.75]{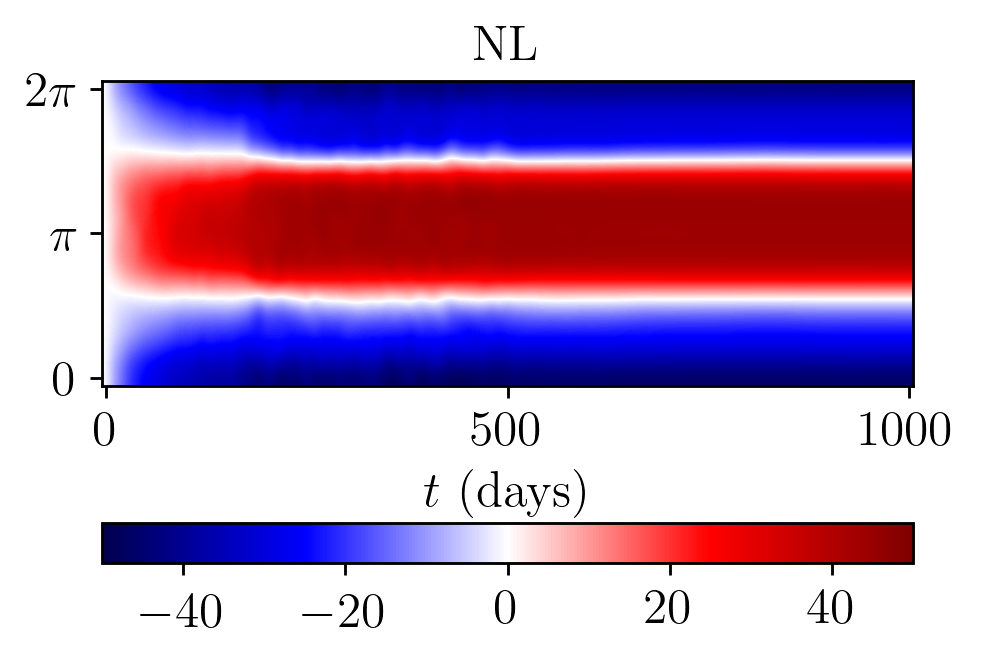}}\\
    {\includegraphics[scale=0.75]{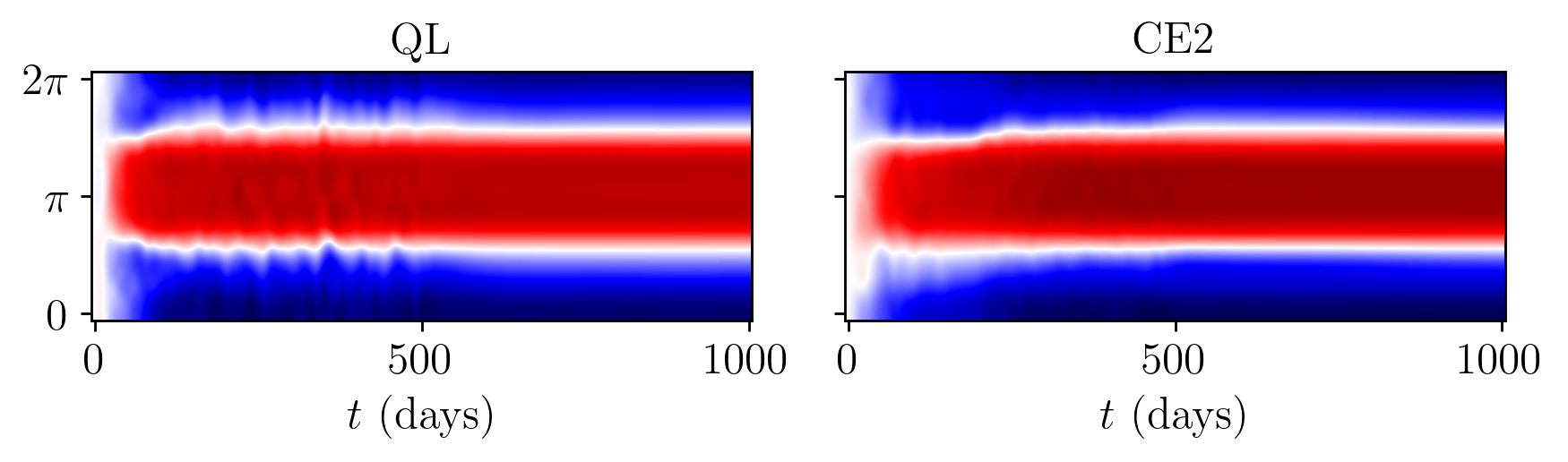}}\\
    {\includegraphics[scale=0.75]{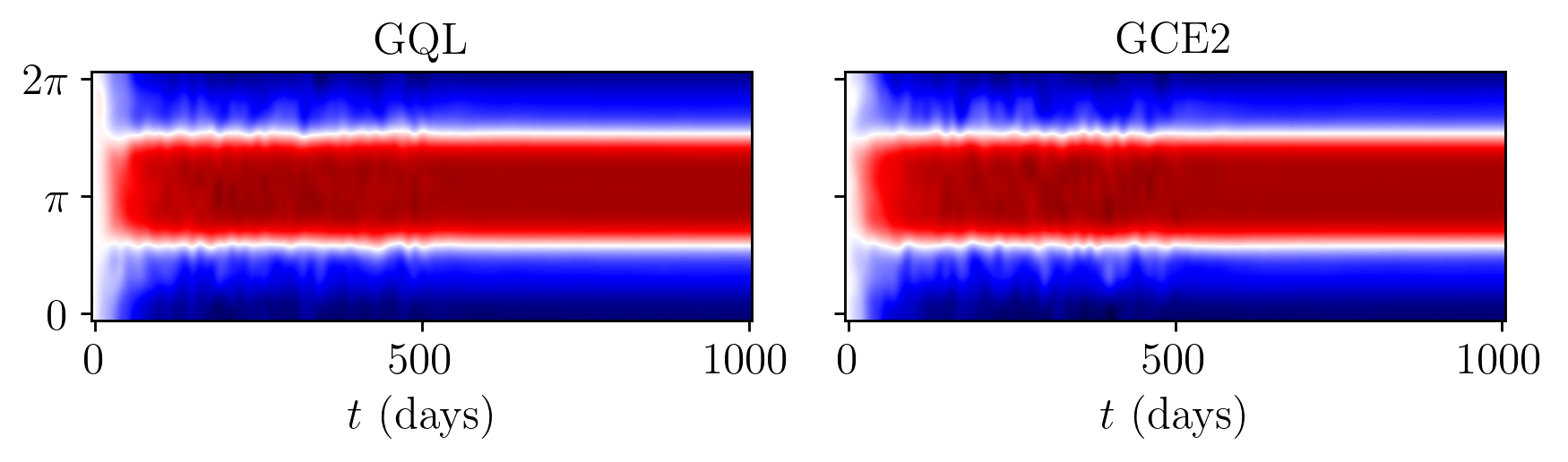}}
    \caption{H\"ovm\"oller  plots showing $\zeta(y,t)$ for NL (top), QL versus CE2 (middle) and GQL versus GCE2 (bottom) for the Kolmogorov flow case. All colour ranges are identical. Time-averaging commences at $t = 500$ days.}
    \label{fig:kf:hoevmoeller}
\end{figure*}

\Cref{fig:kf:hoevmoeller} contains H\"ovm\"oller plots of $\zeta(y,t)$ showing the evolution of the central band of vorticity with time for the five different equation systems. All solutions evolve from the same initially random noise field of power $10^{-4}$. Time-averaging is conducted in the window $500 < t < 1000$ days. Up until time-averaging begins, the QL, GQL and GCE2 systems exhibit noticeable fluctuations in jet location over a relatively small time-scale of $t \sim 10$ days. Such fluctuations appear to be much less pronounced in the NL and CE2 systems, wherein the vorticity jet appears to stabilise at its location at around $t = 250$ days. The time-averaged solution of all five systems are in excellent qualitative agreement, with the possible exception of QL which appears to predict a slightly weaker jet (lighter red colour in the middle left panel).

\begin{figure*}
    \centering
    {\includegraphics[scale=0.65]{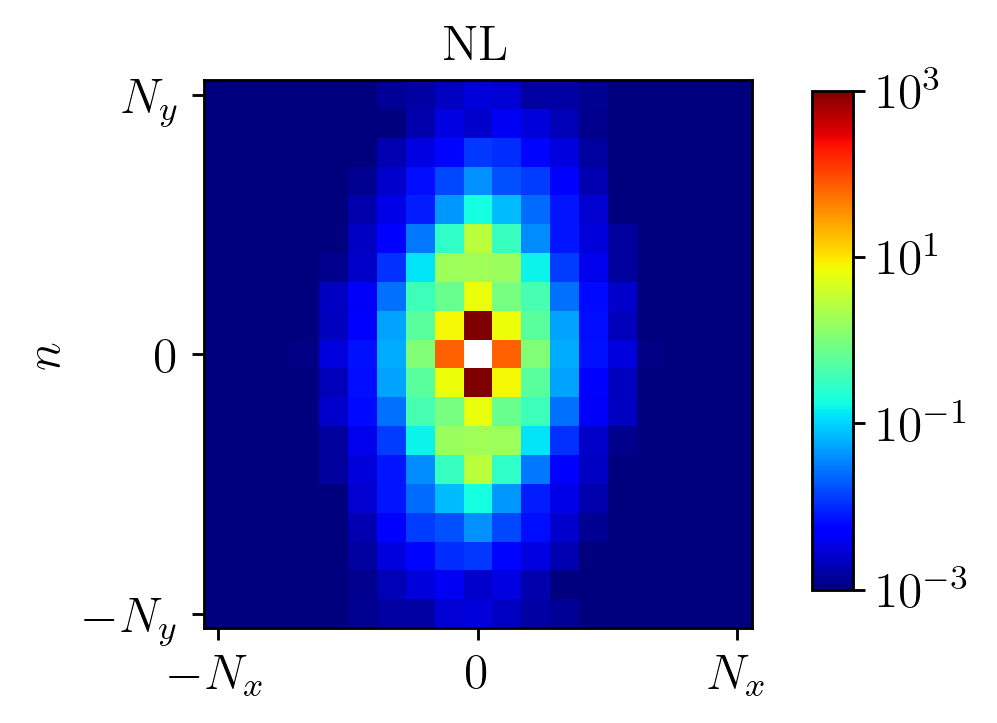}}\\
    {\includegraphics[scale=0.65]{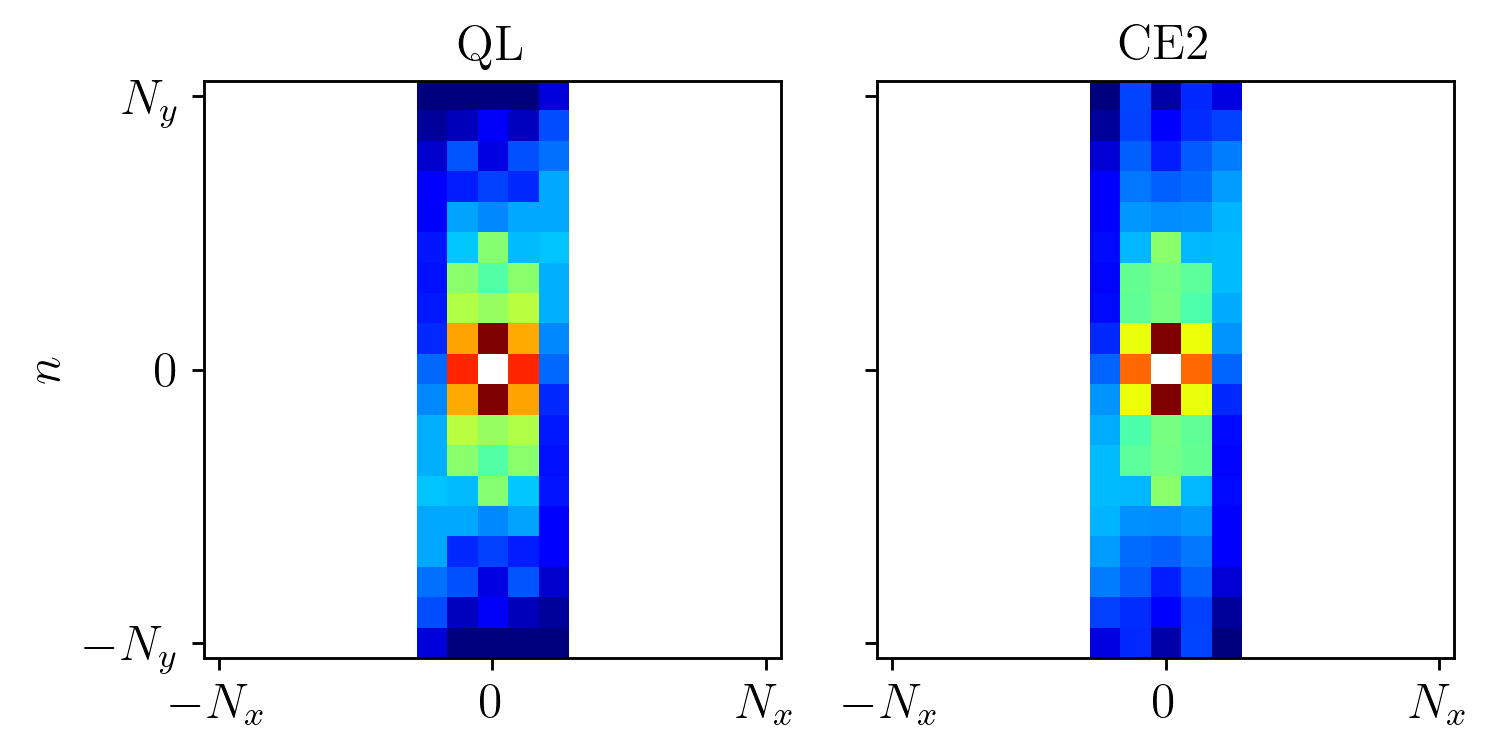}}\\
    {\includegraphics[scale=0.65]{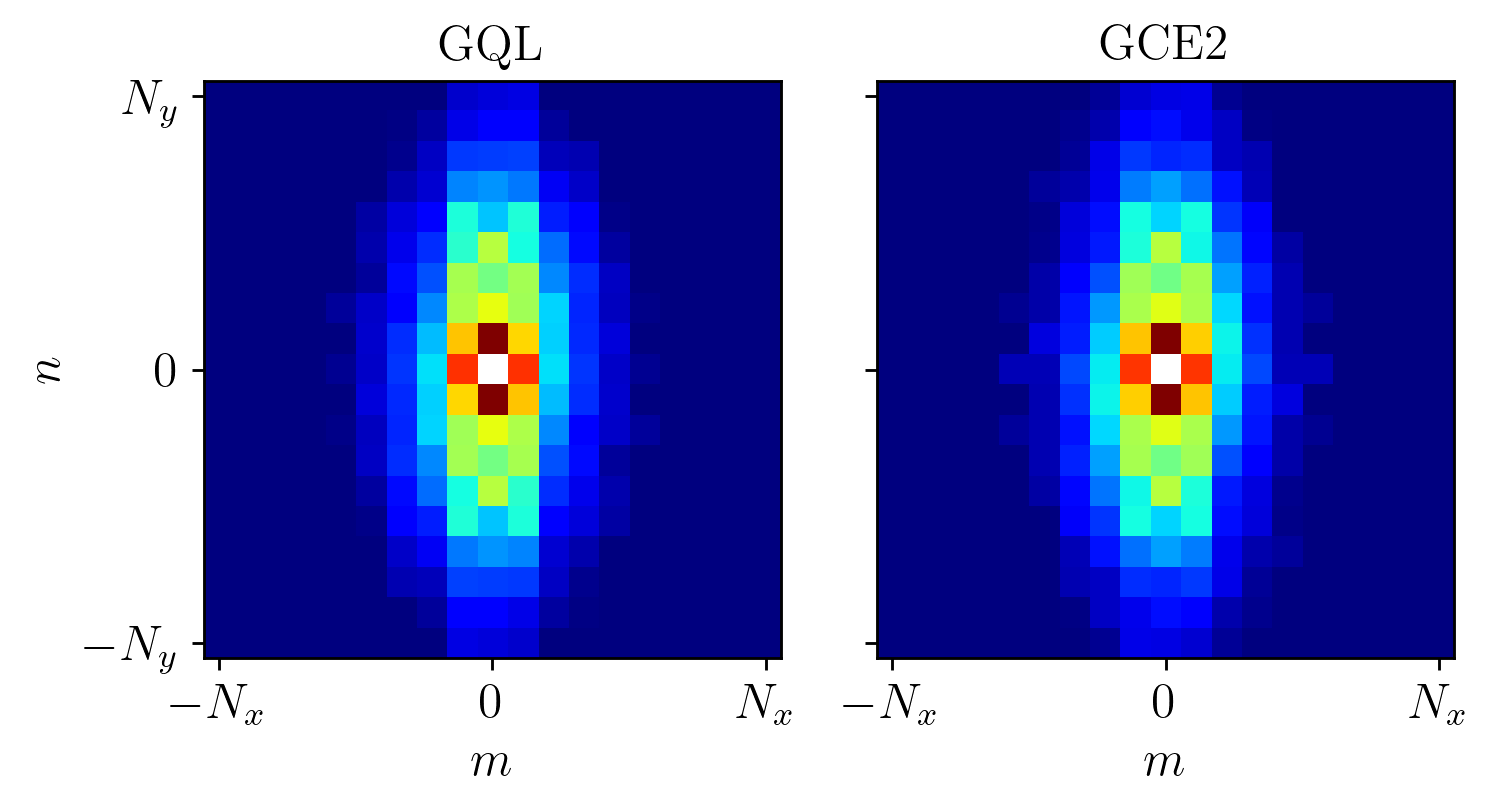}}
    \caption{Time-averaged energy spectra $\overline{E(m,n)}$ at $t = 1000$ days for NL (top), QL versus CE2 (middle) and GQL versus GCE2 (bottom) for the Kolmogorov flow case.}
    \label{fig:kf:spectra}
\end{figure*}

Focusing on subtle quantitative differences, we show time-averaged energy spectra for the various solutions in \cref{fig:kf:spectra}. Each panel plots \NEW{$E(m,n) = \hat\zeta_{m,n}^*\hat\zeta_{m,n}/{\left(m^2 + n^2\right)}$} with zonal wavenumbers $m$ running along the $x$-axis and non-zonal wavenumbers $n$ along the $y$-axis. Evidently, all equation systems predict energy to be primarily concentrated in two pairs of conjugate modes, namely $(m = 0, n = \pm 1)$ and $(m = \pm 1, n = 0)$ (N.B. the mode $(m = 0,~ n = 0)$ denotes the mean across the entire domain and has been set to contain no energy in our solutions). In the fully non-linear case (top panel),  the remaining energy is spread over spectrally-local modes within a relatively narrow band of zonal wavenumbers, with remaining zonal wavenumbers containing small but non-zero energy. In QL and CE2 (middle panels), this zonal spreading of energy is limited to $|m| \le 2$, with no energy in modes with larger zonal wavenumbers. On close inspection, minute differences between QL and CE2 can be observed. For instance, QL (middle left panel) contains slightly higher energy in the $(m = \pm 1, n = 0)$ harmonic pair (dark orange squares) and a slightly different energy distribution across the zonal mean modes $(m = 0,n)$ as compared with CE2 (middle right panel).

The GQL and GCE2 solutions (bottom panels), however, are in excellent agreement with each other. Moreover, they both clearly improve upon the QL/CE2 solution (middle panels) when compared with the NL solution (top panel). Both GQL and GCE2 exhibit similar spreading of energy around the four most energetic modes as seen in NL, albeit with slightly less energy in the outer bands of modes with $|m| = 4,5$. Crucially, the modes $|m| > 5$ contain a small amount of energy in a similar manner to NL, and as opposed to QL/CE2 where energy is completely absent in these modes. These results validate GCE2 as a statistical theory for GQL dynamics, with the ability to improve upon predictions of CE2.

\begin{figure}
    \centering
    \includegraphics[scale=0.6]{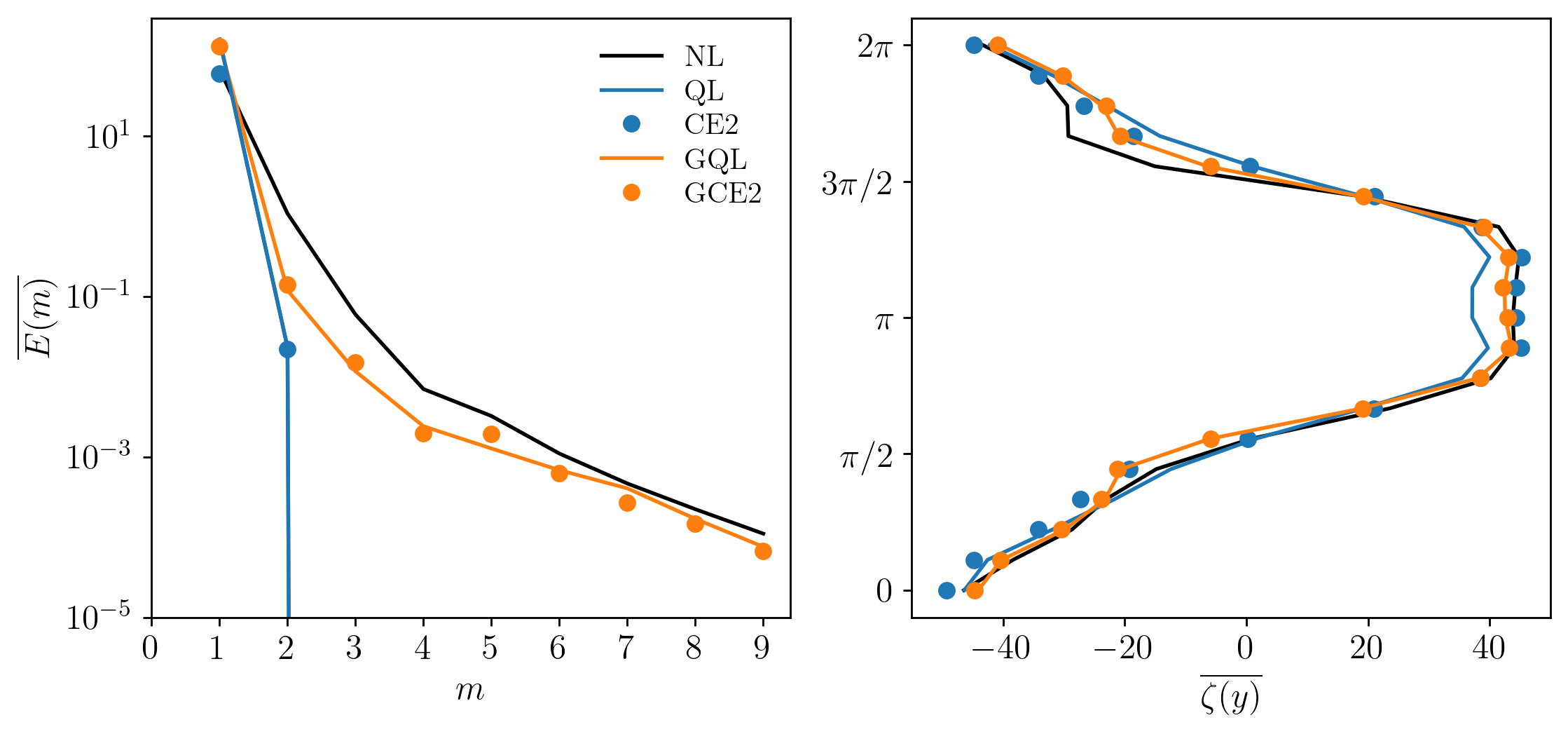}
    \caption{Left: One-dimensional slice of the time-averaged energy spectrum $\overline{E(m,0)}$ predicted by the different equation systems for the Kolmogorov flow case. Right: zonal mean vorticity profile $\zeta(y;t = t_\infty)$}
    \label{fig:1dspectra:kf}
\end{figure}

We now proceed to make quantitative comparisons between the various solutions to elucidate on the qualitative differences observed above. We show the energy distribution over zonal wavenumbers $m$ in the $n = 0$ slice of the time-averaged energy spectra in \cref{fig:1dspectra:kf} (left panel). NL (black line) contains energy in all non-trivial zonal wavenumbers $m \in [1,9]$. Moving away from the strongest wavenumber $m = 1$, energy tapers-off rapidly across the wavenumbers $m \le 4$ and more gently along the wavenumbers $m \ge 5$. GQL (orange line) and GCE2 (orange dots) mimic this distribution of energy reasonably well; however, the $m = 1$ mode is slightly stronger and the wavenumbers $m \le 4$ have slightly lower energy in GQL/GCE2 than NL. Interestingly, minor departures of energy between GCE2 and GQL are apparent, particularly for $m = 5,7$. In comparison, QL (blue line) and CE2 (blue dots) contain most energy in two wavenumbers. Comparing these modes carefully, we note that QL contains significantly more energy in the $m = 1$ mode than does CE2, showing that QL and CE2 solutions also have disagreements.

In the right panel of \cref{fig:1dspectra:kf}, we plot the zonal mean vorticity as a function of the $y$-coordinate. The centreline vorticity ($y = \pi$) of the NL solution is quite well predicted by CE2, followed closely by GQL and GCE2. However, the QL solution departs conspicuously from the remaining predictions at the centreline. Differences between the solutions away from the centreline are relatively small, although, at $y \sim 3\pi/2$, the NL solution is over-predicted by both solution pairs GQL/GCE2 and QL/CE2. Again, the QL and CE2 solutions exhibit significant departures from each other. Though we have noted here a difference between QL and CE2 solutions, no major departures are immediately evident between the GQL and GCE2 solutions. 

The results above demonstrate that GQL/GCE2 improve upon the predictions of QL/CE2 when compared with the NL solution. Barring the minor differences, these results also validate the predictions of DSS (CE2 or GCE2) against the corresponding dynamical systems (QL or GQL, respectively). Now we focus our attention on investigating the origin of divergences between each solution pair QL/CE2 and GQL/GCE2. As mentioned earlier, \cite{Nivarti2022} pointed out recently that divergences can and do arise between the QL and CE2 solutions despite the mathematical correspondence between them. Importantly, they linked such divergences to a rank instability available to the dynamics of the CE2 system and prohibited within the QL system. Divergences occur when the rank of a zonal submatrix $C^{(m)}$ in the CE2 second cumulant departs from its initial unity value -- the corresponding rank in QL must always remain unity, thus constituting an important source of differences. \NEW{We define the matrix $C^{(m)}$ to be the zonal decomposition of the second cumulant: $C^{(m)} = \hat{c}_{\zeta \zeta} (m, n_1; m, n_2) = \frac{1}{(2\pi)^3}\int_{0}^{2\pi}\int_{0}^{2\pi}\int_{0}^{2\pi}c_{\zeta\zeta}(x_1,y_1; x_2, y_2) e^{i m (x_2 - x_1) - i n_1 y_1 - in_2 y_2} dx_1 dx_2 dy_1 dy_2$ for $m \in [1,M)$ where $M$ is the spectral cutoff in the zonal wavenumber.} In order to compare GQL against GCE2, the zonal projection of the GCE2 field  can be employed.

\begin{figure*}
    \centering
    {\includegraphics[scale=0.7]{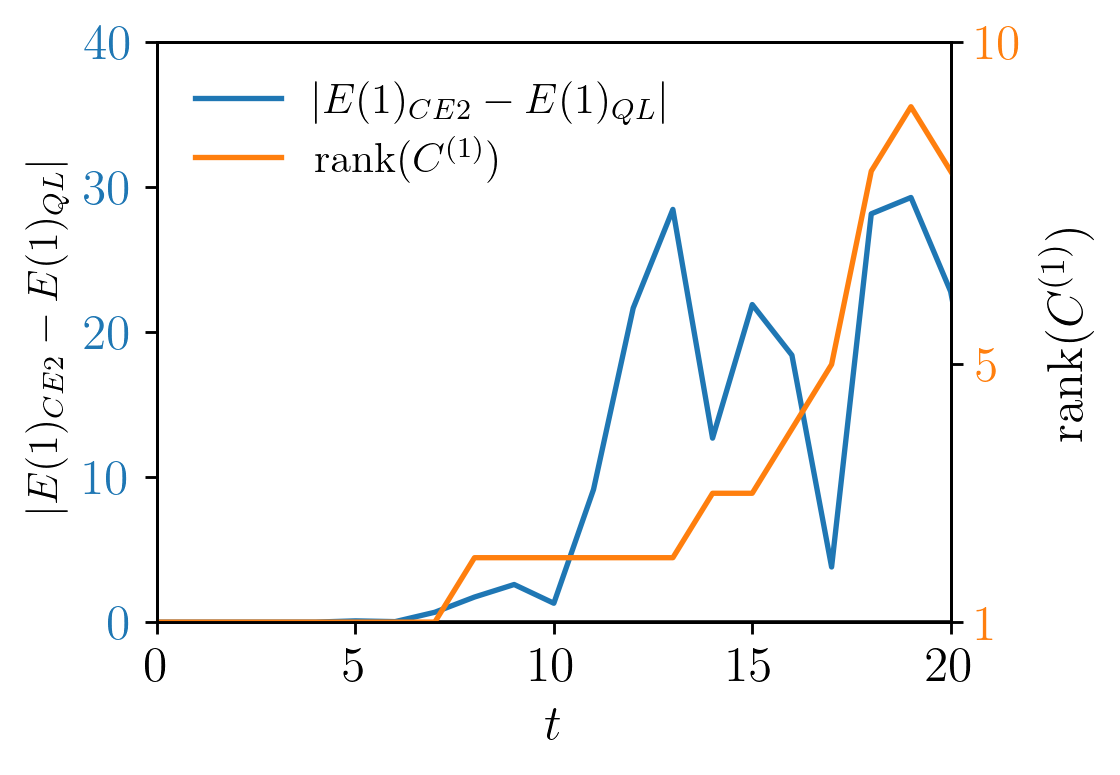}}\\
    {\includegraphics[scale=0.7]{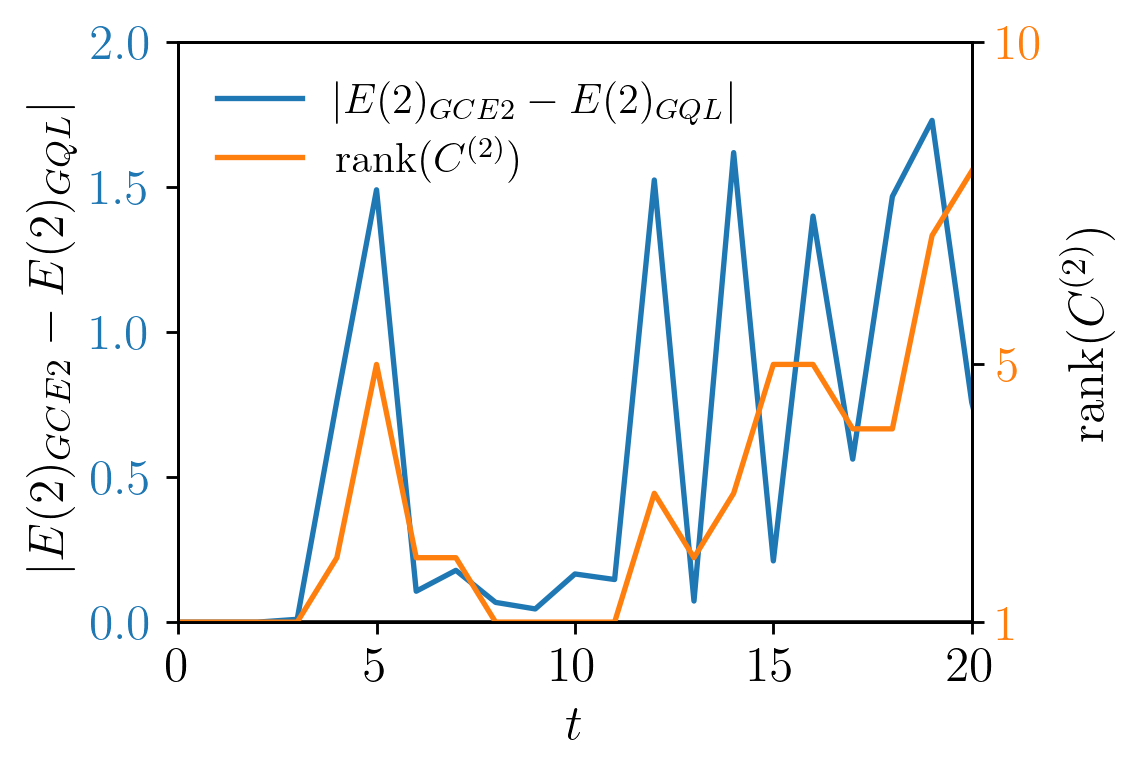}}
    \caption{Difference in energy $E(m)$ of zonal mode $m$ for the QL and CE2 solutions ($m = 1$, top) and for the GQL and GCE2 solutions ($m = 2$, bottom), with the corresponding rank $C^{(m)}$ shown on the right. The first $20$ days of the Kolmogorov flow case are shown with half the timestep size as before. In each comparison, the divergence of zonal energies in the dynamical (QL, GQL) and statistical (CE2, GCE2) solutions appears to be strongly associated with the onset of rank instability in the latter.}
    \label{fig:kf:rank}
\end{figure*}

{\Cref{fig:kf:rank} simultaneously plots the time-evolution of two different quantities for $0 \le t \le 20$ days. The left vertical axis quantifies the absolute difference in zonal energy $|E(m)_{CE2} - E(m)_{QL}|$ between CE2 and QL for a given zonal mode $m$ plotted using blue lines. The right vertical axis quantifies the rank $\textrm{rank}(C^{(m)})$ of the cumulant submatrix $C^{(m)}$. \NEW{$E(m) = \sum_n \hat{\zeta}_{m,n}^*\hat{\zeta}_{m,n}/(m^2 + n^2)$ where $\hat{\zeta}_{m,n}$ is the relevant Fourier coefficient of vorticity $\zeta(x,y)$, which is calculated in QL and GQL as $\zeta = {\zeta}_\ell+ \zeta_h$. In CE2 and GCE2, $E(m)$ can be calculated as $E(m) = \sum_n |\hat{c}_\zeta(n)|^2/n^2$ for $m = 0$ and $E(m) = \sum_n C^{(m)}(n,n) /(m^2 + n^2)$ for $m \ge 1$.} The top panel in} \cref{fig:kf:rank} compares the QL and CE2 solutions considering the zonal mode $m = 1$. As both systems evolve from an identical random noise initialisation, the difference $|E(1)_{CE2} - E(1)_{QL}|$ (blue line) is zero initially. The zonal energies begin to differ around $t = 5$ days showing an increasing magnitude of differences with time. Simultaneously with the emergence of differences, the rank of second cumulant submatrix $C^{(1)}$ in CE2 (orange line) departs from unity, increasing nearly-monotonically until $t = 20$ days. This is a clear indication of the strong link between the onset of rank instability in CE2 and CE2's divergence from QL as also found by \cite{Nivarti2022}. In a similar manner, the bottom panel in \cref{fig:kf:rank} compares the same quantities for GQL and GCE2 with spectral cutoff $\Lambda = 1$ for the high mode $m = 2$. The zonal energy difference $|E(2)_{GCE2} - E(2)_{GQL}|$ (blue line) shows that the respective solutions begin to diverge even before $t = 5$ days. Again, this happens simultaneously with the departure of the corresponding submatrix $C^{(2)}$ (orange line) rank from unity. In fact, in this case, we also observe that at times when the solutions have very similar energy (say at $t \sim 8$ days), the rank also returns to unity. These results confirm that, akin to divergences observed previously for QL and CE2 \citep{Nivarti2022}, predictions of GQL and GCE2 may also diverge from one another as a result of the onset of rank instabilities.  \NEW{We have reproduced the GCE2 rank instability using independently written code (see \cite{mt_2023}) for the case of Kolmogorov forcing on the sphere.}

\begin{figure*}
    \centering
    {\includegraphics[scale=0.75]{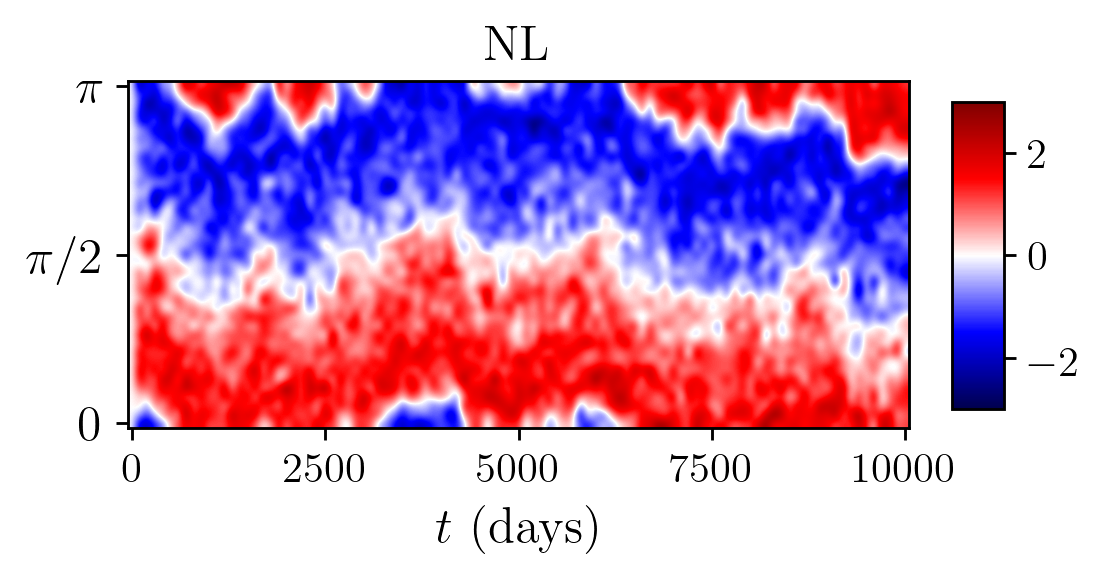}}\\
    {\includegraphics[scale=0.85]{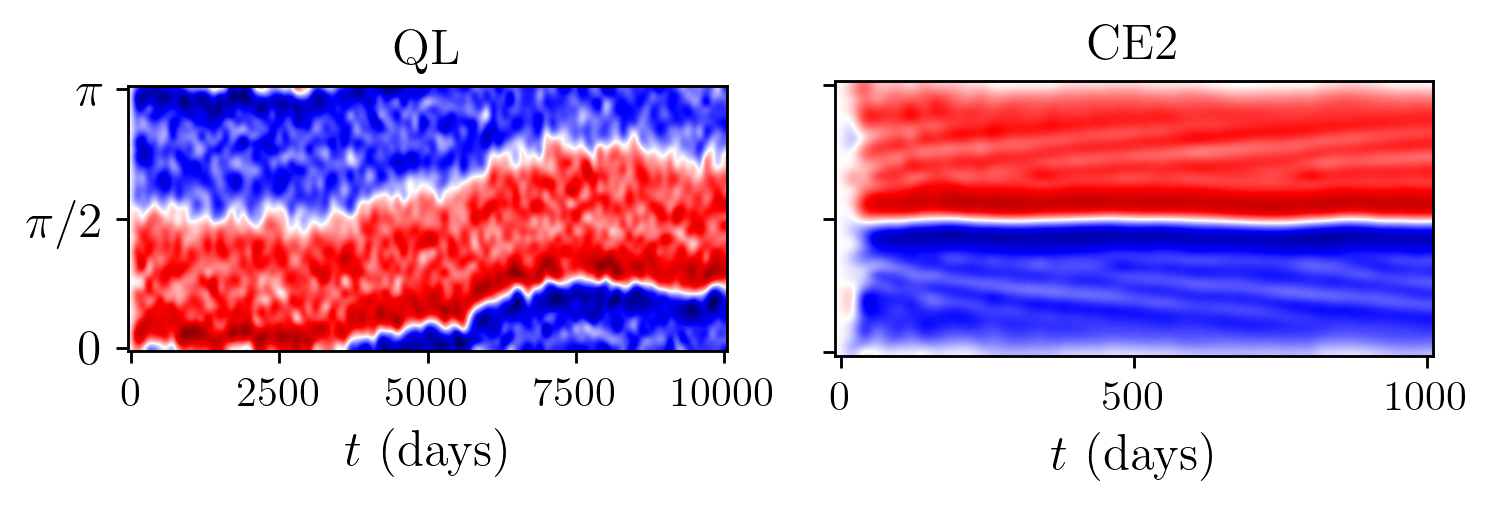}}\\
    {\includegraphics[scale=0.85]{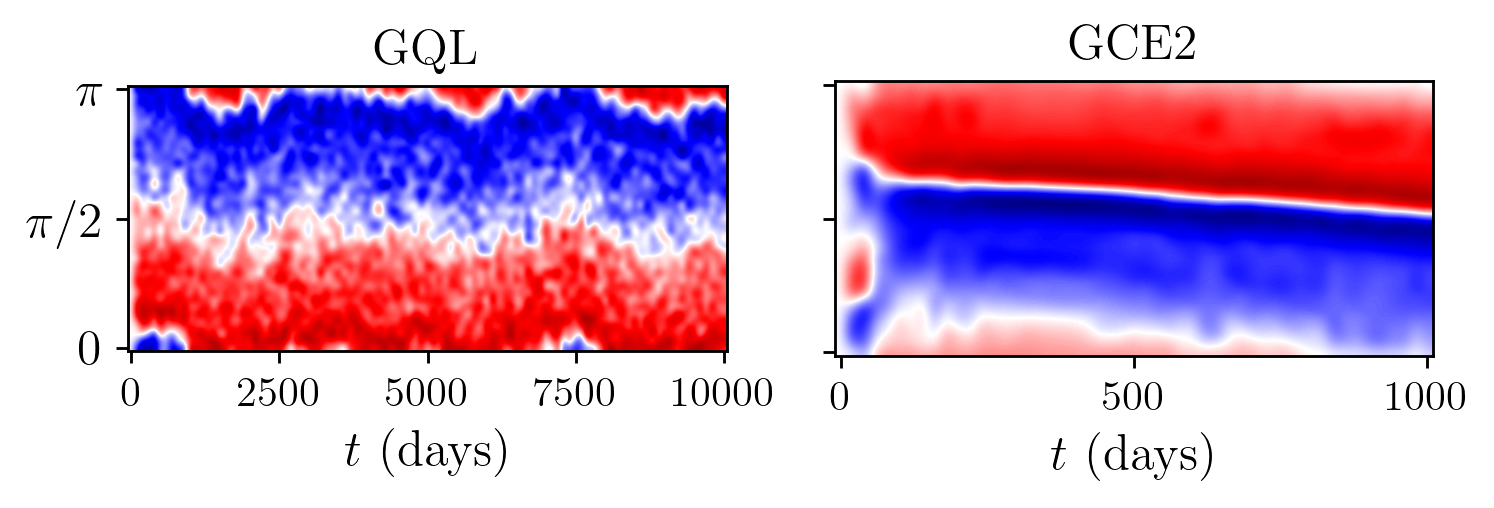}}
    \caption{H\"ovm\"oller  plots showing $\zeta(y,t)$ for NL (top), QL versus CE2 (middle) and GQL versus GCE2 (bottom) for a the stochastically forced case with resolution $M = 12, N = 20$. All colour ranges are identical. Jet migration is captured by GCE2 whereas CE2 fails to capture it.}
    \label{fig:sf:hoevmoeller}
\end{figure*}
\subsection{Narrow-band stochastic forcing}
We now consider a stochastically-driven system on the rotating $\beta$-plane. We adopt the formalism detailed in  \cite{Constantinou2016} which was used for the validation of the Statistical State Dynamics (SSD) model. The forcing term $F$ in the vorticity equation becomes
\begin{equation}
    F(\bm{x},t) = \NEW{\sqrt{\varepsilon} \xi},
    \label{eq:sf}
\end{equation}
\NEW{where $\varepsilon$ is the energy injection rate and ${\xi}$ is the white-in-time stochastic noise with zero mean and covariance $Q(\bm{x})$. Following \cite{Constantinou2015} (see H.4 there), the covariance $Q(\bm{x})$ is specified on the Fourier domain by the forcing spectrum $\widehat{Q} (\bm{k}) = c(m)^2 d^2 e^{-n^2d^2}$ which is non-zero for zonal wavenumbers $m \in [m_f, m_f + \delta m)$ where $d = 0.1$ and $c(m)$ is such that the total contribution of $\widehat{Q}(\bm{k})$ is unity for each zonal wavenumber \cite[see H.5]{Constantinou2015}, and as a result the total energy injection rate of $F(\bm{x},t)$ in \cref{eq:sf} is $\varepsilon$. Here, we have set $m_f = 8$ and $\delta m = 2$ (two zonal wavenumbers $8, 9$ are forced) on a $12\times20$ grid.} \NEW{Because stochastically-driven modes have zonal wavenumbers that exceed the cutoff $\Lambda$ the covariance of the stochastic forcing is added to the tendency equation for the second cumulant, Eq. \ref{eq:gce2:c2}.}


\Cref{fig:sf:hoevmoeller} shows H\"ovm\"oller plots $\zeta(y,t)$ for a case corresponding to the equatorial $\beta$-plane ($\theta = 0^\circ$) with $\beta = 10.0$ and $\mu = 0.01$. As is customary \citep{Tobias2013}, a hyperviscosity coefficient $\nu_4$ is applied such that the largest wavenumber dissipates energy at unit rate. The energy injection rate (described above) via the forced zonal wavenumbers is $\varepsilon = 0.02$. The dynamical equations (NL,QL and GQL) are solved for $10000$ days, whereas the statistical equations (CE2 and GCE2) need only be solved for a tenth of that time.

The NL solution (top row in \cref{fig:sf:hoevmoeller}) consists of two opposed jets of vorticity that migrate gently downwards beginning at $t = 5000$ days. The migration appears to occur at constant speed (fixed slope in $t-y$ coordinates) {as has been identified previously by \cite{Cope2020}}. The QL solution (middle left panel) predicts instead a relocation of the positive vorticity jet from the lower half of the domain to its centreline during the period $2500 < t < 7500$ days. After this period, the jet stabilises to a steady position. No vertical movement is recorded in the CE2 solution (middle right panel), which merely captures the dual-jet solution with the sharp inter-jet boundary lying along $y = 0 \forall t \in [0,1000]$ so the migration is not captured by a QL theory, as also determined by \cite{Cope2020}. The GQL and GCE2 solutions are shown in the bottom row. Whereas GQL (bottom left) appears to predict a mild degree of jet migration, GCE2 (bottom right) records the downward jet migration as observed in NL (top)! Note that the slope in GCE2 must be magnified ten times for an apples-to-apples comparison with that in NL due to the former's much shorter run. Here too then, GCE2 improves significantly over CE2 by capturing jet migration that CE2 can not.

\begin{figure}
    \centering
    \includegraphics[scale=0.84]{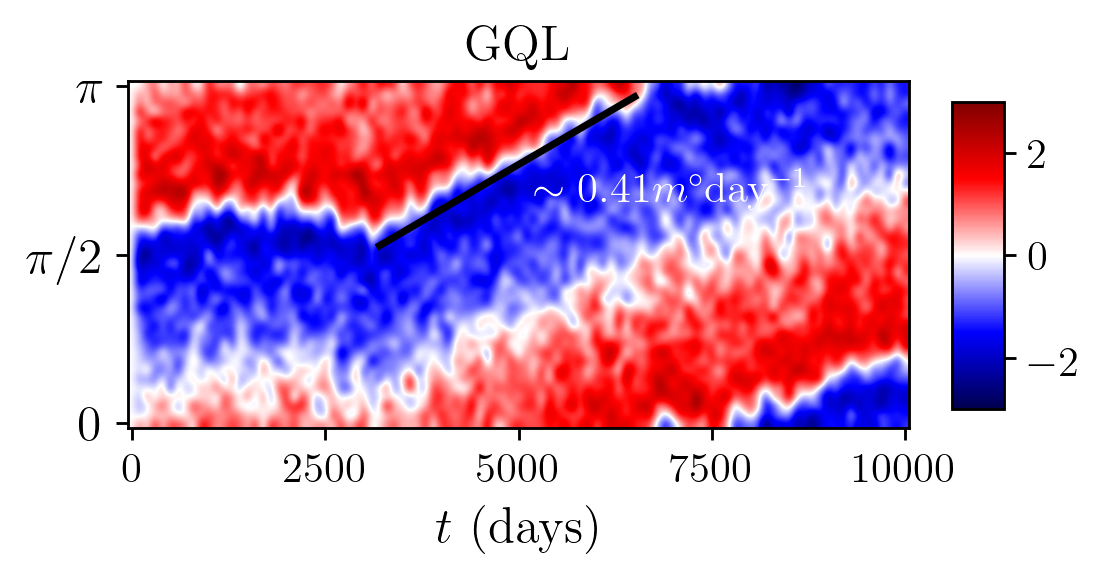} \includegraphics[scale=0.745]{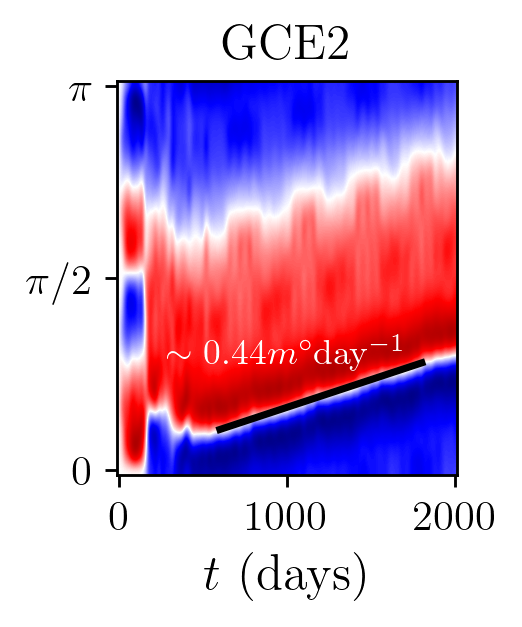}
    \caption{Left: GQL instance from of an ensemble of differently-seeded random noise ICs. Right: GCE2 initialised with a maximum ignorance initial condition run for a shorter period (colour range is identical to left figure). GCE2 with maximum ignorance captures jet migration {with a similar speed (indicated in $m^\circ = 10^{-3}$ degrees per day)} as seen within a large ensemble of GQL runs.}
    \label{fig:sf:ensemble}
\end{figure}

%
\begin{figure*}
    \centering
    {\includegraphics[scale=0.7]{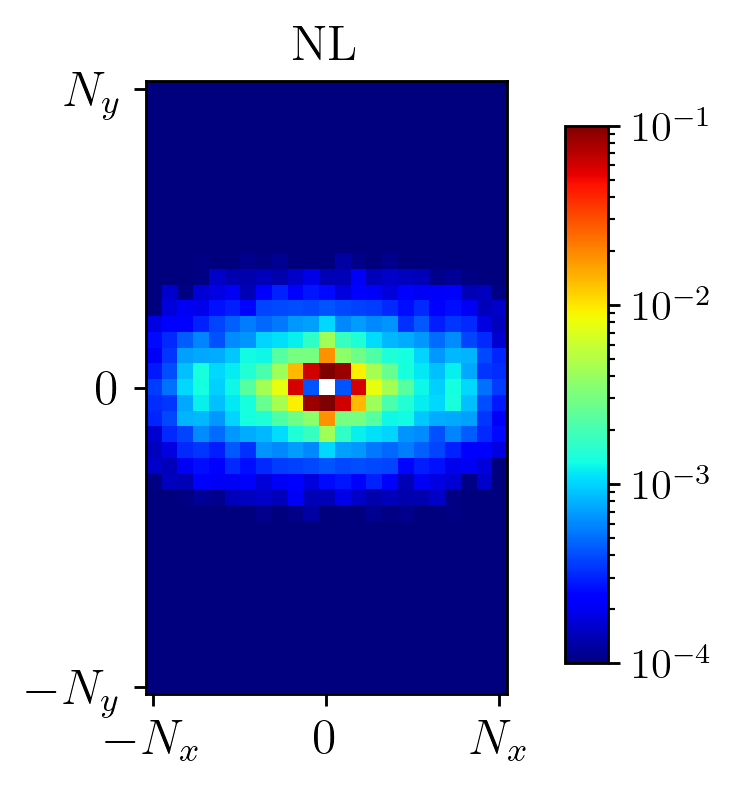}}\\
    {\includegraphics[scale=0.7]{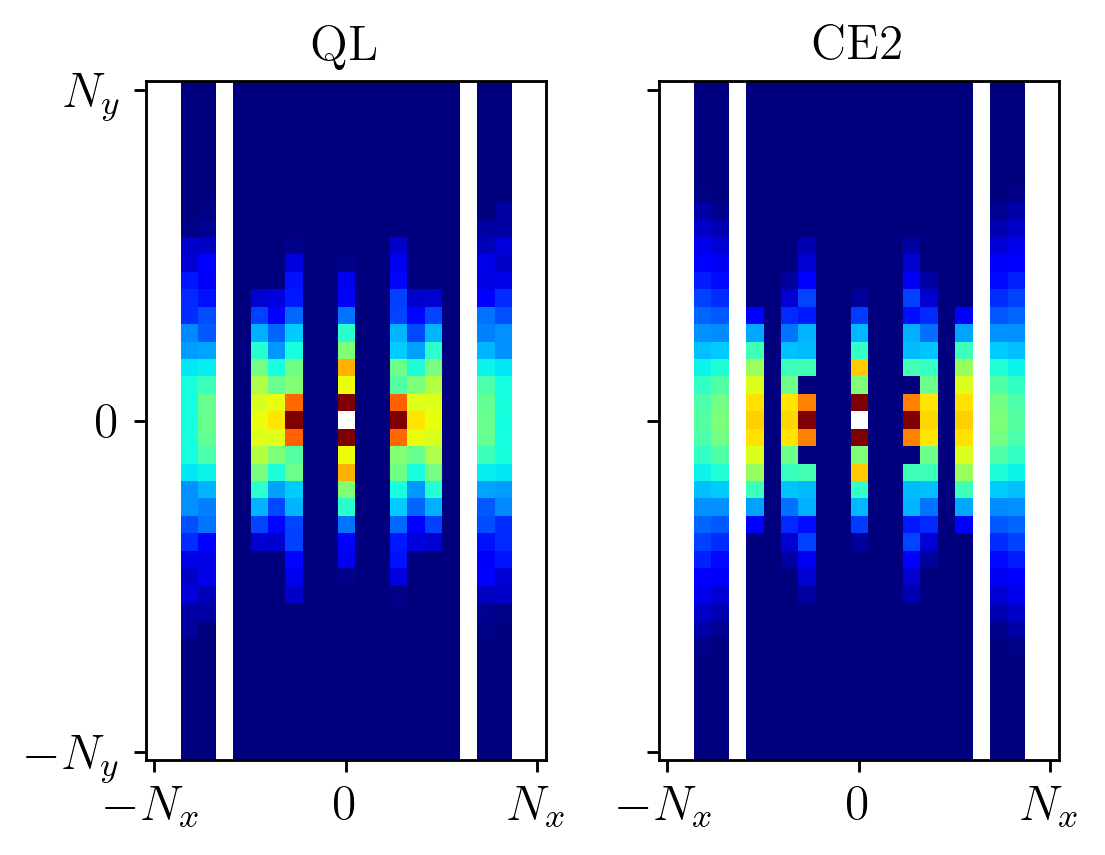}}\\
    {\includegraphics[scale=0.7]{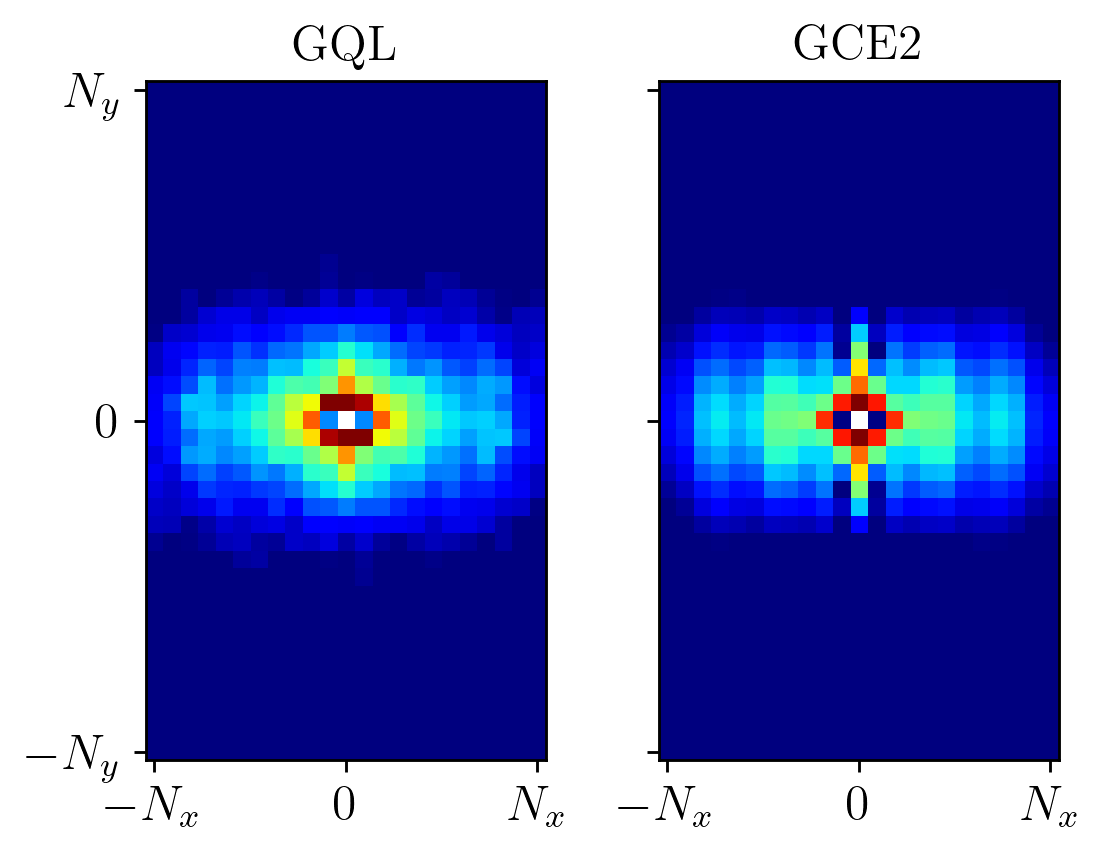}}
    \caption{Time-averaged energy spectra $\overline{E(m,n)}$ for NL (top), QL versus CE2 (middle) and GQL versus GCE2 (bottom) for the stochastically forced case. GCE2 improves considerably over CE2, but also diverges from GQL.}
    \label{fig:sf:spectra}
\end{figure*}
In order to elucidate on the differences in jet behaviour recorded by GQL and GCE2, we conducted a series of GQL simulations with different random seed values used to generate the initial conditions. This resulted in an ensemble containing a wide range of jet behaviour, including jet migration and steady jet propagation. In the interest of brevity, we show in \cref{fig:sf:ensemble} the results of a single GQL run (left) chosen from the ensemble that predicts upward jet movement {with a speed of roughly $0.41m^\circ$ per day}. We compared this with a GCE2 simulation (left) initialised with maximum ignorance, i.e. with a full-rank second cumulant initialised with power $10^{-6}$. GCE2 records the identical upward jet migration {with a speed of roughly $0.44m^\circ$ per day. Note that the x-axis ranges in both figures are significantly different, so the slopes do not appear identical visually.} It is interesting that GCE2 with a maximum ignorance initial condition predicts upward jet migration rather than in the opposite direction {as observed in GCE2 with a unity rank IC (see \cref{fig:sf:hoevmoeller})}.

In \cref{fig:sf:spectra}, we show time-averaged energy spectra for the different solution methods for the stochastically-forced case shown in \cref{fig:sf:hoevmoeller}. The fully-nonlinear system (top panel) contains energy in the $m = 1$ zonal mode, primarily via the $n = \pm 1$ non-zonal wavenumber (which corresponds to the opposed jet configuration). A significant portion of the energy is also present in the $m = 1$ and $m = 2$ zonal modes, the remaining being scattered through the entire range of zonal modes but a relatively narrow band of non-zonal modes. In comparison with NL, the QL and CE2 solutions (middle row) consist of localised bands of excited zonal modes. For instance, the $m = 1$ and $m = 2$ modes are relatively very weak in the QL and CE2 solutions; so also, the modes $m = 6$ or $m \ge 9$ appear to be absent in comparison to NL. This localisation of energy arises because scattering is unavailable in QL (and therefore CE2) owing to the absence of the required non-linearities; a given zonal mode $m$ becomes energetic in QL only via a corresponding instability of the mean flow. Once energetic, the zonal mode $m$ can transfer its energy back to the mean via self-interactions, but may not transfer energy to another non-mean mode $|m| \ne 0$. On closer examination, it is revealed that QL and CE2 exhibit differences. The differences between energy distribution over non-zonal wavenumbers in the $m = 2$ mode are immediately evident between QL (left panel) and CE2 (right panel). {In other words, the CE2 solution obtained \emph{is not identical} to that of a single realisation of QL.} As in the Kolmogorov flow case, we hold that these divergences are linked to the rank instability in CE2.

\begin{figure}
    \centering
    \includegraphics[scale=0.6]{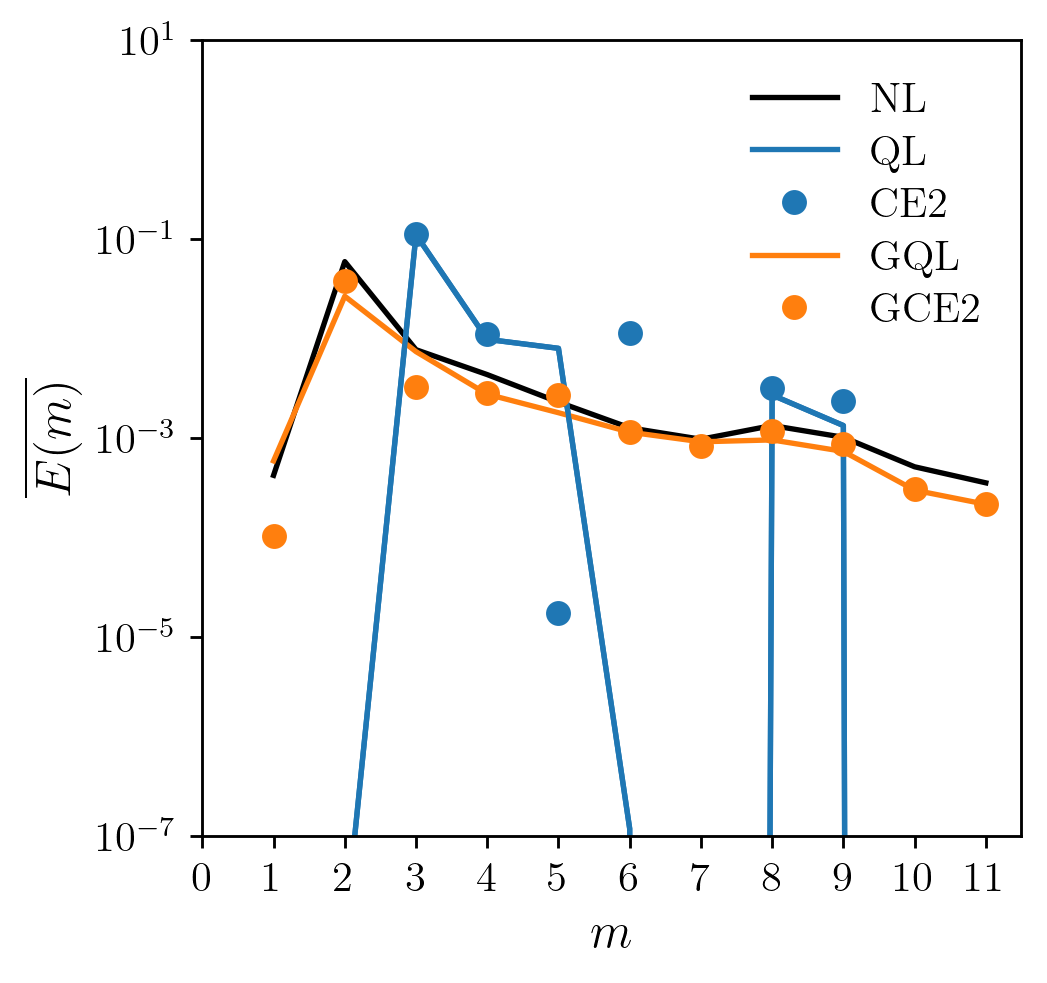}
    \caption{One-dimensional slice of the time-averaged energy spectrum $\overline{E(m,0)}$ predicted by the different equation systems for the stochastically forced case.}
    \label{fig:1dspectra:sf}
\end{figure}

The GQL and GCE2 solutions in \cref{fig:sf:spectra} (bottom row) improve considerably over the QL/CE2 solutions. The missing zonal modes $m = 1,2,6,$ etc. are found to be energetic in GQL/GCE2. This is because non-linear interactions involving the $m = 1$ mode that are available within GQL (and thereby GCE2) allow for scattering of energy leading to a broader spread of energy over the range of zonal wavenumbers. It is remarkable though that GQL with a cutoff $\Lambda = 1$ appears to be in excellent agreement with NL (top panel). We note that GCE2 diverges from GQL to some extent in the same way CE2 was seen to diverge from the QL solution. Broadly speaking, energy appears to be less spread out in GCE2 as compared to GQL, being more concentrated to small non-zonal wavenumbers. For instance, energy in the $m = 1$ mode in GCE2 (bottom right panel) is limited to a significantly narrower range of non-zonal wavenumbers $|n| \lesssim 3$. Additionally, the $(\pm 1,0)$ modes are weaker in GCE2 by an order of magnitude.

\begin{figure*}
    \centering
    {\includegraphics[scale=0.75]{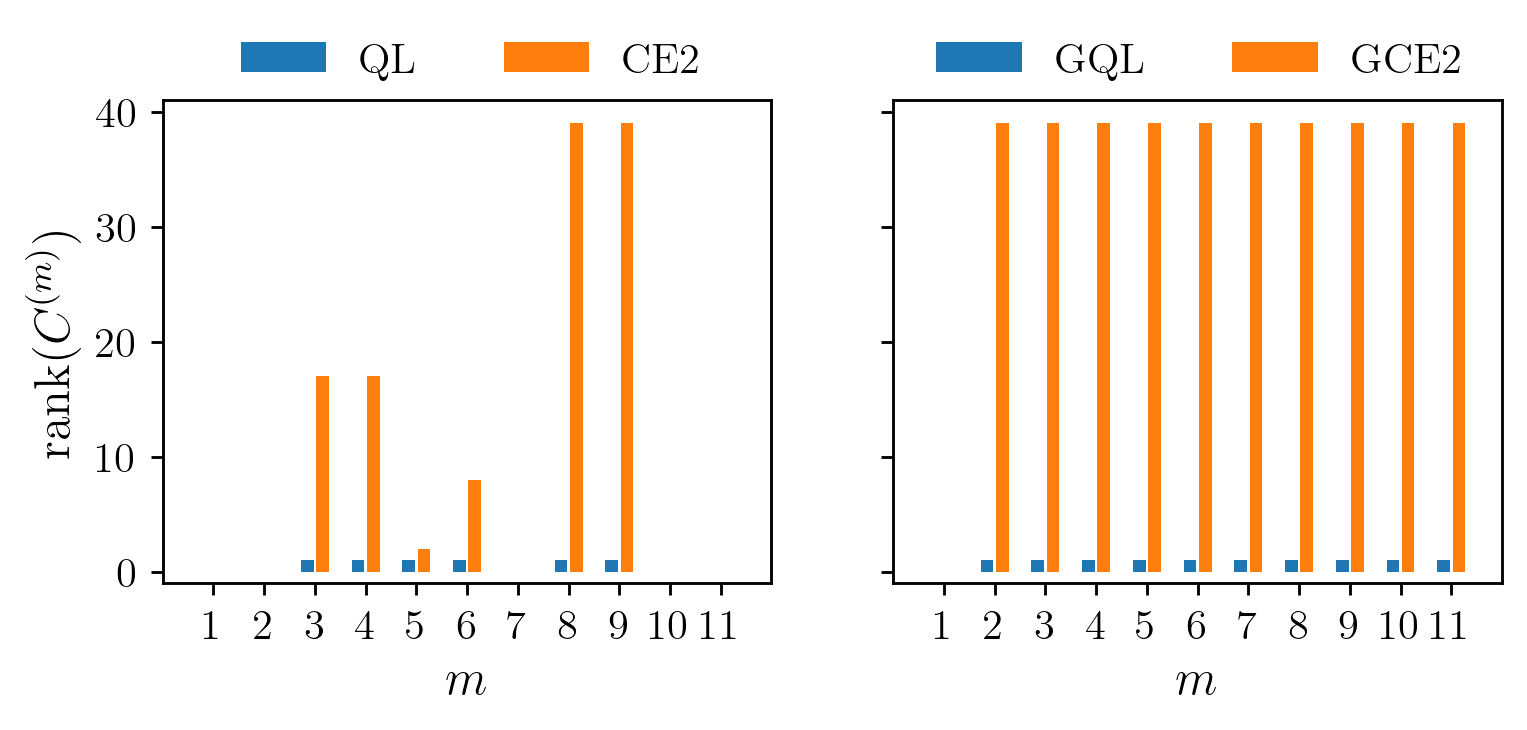}}
    \caption{Comparisons of ranks $C^{(m)}$ in the end point solution for QL and CE2 (left) and for GQL and GCE2 with $\Lambda = 1$ (right). In CE2, each zonal mode undergoes its own rank instability, whereas in GCE2, the allowed $HL\rightarrow H$ interactions cause ``rank-scattering'', and the (full) rank of forced zonal modes ($m = 8,9$) is adopted by all other high modes.}
    \label{fig:sf:ranks}
\end{figure*}

\NEW{We show the energy distribution over zonal wavenumbers $m$ in the $n = 0$ slice of the time-averaged energy spectra for the stochastically forced case in \cref{fig:1dspectra:sf}. NL (black line) is shown to contain energy in all non-trivial zonal wavenumbers $m \in [1,11]$ supporting the findings of \cref{fig:sf:spectra}. However, \cref{fig:1dspectra:sf} clarifies that the $m = 1$ and $m = 2$ zonal wavenumbers are respectively associated with the lowest and highest energies. Moving away from the wavenumber $m = 2$, energy tapers-off steadily across the remaining wavenumbers $m \le 11$. The energy distributions predicted by GQL (orange line) and GCE2 (orange dots) are in good agreement with the NL energy distribution, save for the $m = 1$ mode estimated to be slightly weaker by GCE2. At this wavenumber and at $m = 3$, GQL exhibits a slight departure from GCE2, aligning more favourably with NL than GCE2. Contrasting with the GQL and GCE2 solutions, QL (blue line) and CE2 (blue dots) do not contain energy in at least five of the eleven wavenumbers $m \in [1,11]$. CE2 does not contain energy in the zonal wavenumbers $m = {1,2,7,10,11}$.  In comparison, QL is additionally deprived of energy in the $m = 6$ zonal wavenumber also, for which CE2 predicts an energy level comparable to NL/GQL/GCE2. Thus, the wavenumber $m = 6$ is a point of significant departure between the QL and CE2 solutions. The QL and CE2 energy distributions also exhibit a conspicuous departure in the $m = 5$ wavenumber and a relatively minor departure in the $m = 9$ wavenumber. These quantitative comparisons provide affirmation to the following: (1) GQL and GCE2 solutions better estimate the NL energy distribution, and (2) neither pair of dynamical and statistical solutions -- GQL/GCE2 and QL/CE2 -- is devoid of divergence.}


%
%
In light of the divergences shown previously for Kolmogorov forcing, it would suffice here to show a departure in the rank of a second cumulant submatrix to explain the divergences seen above for the stochastically driven case. In \cref{fig:sf:ranks}, we show the final-time ranks of second cumulant submatrices. For CE2 (orange bars in the left panel), the zonal modes $m = 8,9$ are full-rank by virtue of the stochastic driving; however, we note that a number of additional zonal mode submatrices also depart in rank from unity as is held by the corresponding QL solution (blue bars in the left panel). Since there is no pathway of interactions that can transfer energy between the various non-mean zonal modes in QL (and CE2 by extension), each of the zonal modes with non-unity rank must have acquired its own rank instability. This contrasts with GCE2 (orange bars in the right panel) where all high zonal modes $|m| > 1$ for the spectral cuttoff $\Lambda = 1$ are full rank. In essence, scattering of energy allowed within the GQL/GCE2 formalism causes full rank of stochastically-driven modes to be transferred to all remaining zonal modes. {We term this \emph{rank scattering}.} {Thus we observe that communication matters: moving from CE2 to GCE2 changes significantly the channels of communication so that energy scattering and rank scattering can occur.}

\section{Conclusions}

In this paper we have tested a method of Direct Statistical Simulation (DSS) that is obtained as a mathematically exact closure for the Generalised Quasilinear (GQL) equations. This method of DSS, which we term GCE2, adopts generalised cumulant expansions and is capable of systematically interpolating between statistics corresponding to quasilinear and fully non-linear equations. We have implemented GCE2 in a numerical code for simulations on the $\beta$-plane with two driving models, deterministic and stochastic. Our simulations, which employ a minimal  spectral cutoff $\Lambda = 1$,  confirm that GCE2 improves considerably over CE2, the DSS method corresponding to quasilinear dynamics. 

\NEW{We comment here that GQL (and hence its statistical manifestation GCE2) can be derived using the method of multiple scales in space and time (see e.g. \mbox{\cite{mct2016}}). The approximation is made formally rigorous by separating variables into those that undergo rapid dynamics on small scales and more slow dynamics of the large-scale variables. Intriguingly, similar methods have been used for homogeneous, isotropic turbulence \citep[see e.g.]{Laval2001}. That paper showed that the degree of intermittency was crucially determined by the form of the interactions that were included in the model and a new model of turbulence was suggested where  nonlocal interactions coupled large and small scales nonlinearly (with the addition of noise) and the local interactions could be modelled by a turbulent viscosity; it would be interesting to test the applicability of generalisations of such a model to anisotropic flows.}

We have also shown that statistics of GCE2 may depart from those of GQL due to the rank instability as found recently for QL and CE2 \citep{Nivarti2022}.  CE2 and QL solutions may diverge at identical parameters despite the fact that CE2 is an exact mathematical closure for QL. \NEW{Such divergences {(also observed by \cite{Oishi2022})} are linked to the emergence of an instability, the rank instability, that is available within the CE2 system but unavailable in QL. The origin of the rank instability can be understood analytically with a simple linear model \citep{Nivarti2022}.  We have found that the GCE2 system admits such instabilities too, and therefore its solutions can and do divergence from solutions of the GQL system. This is a feature rather than a bug. CE2 is a statistical description, whereas any single realisation of QL dynamics is a dynamical one.  Likewise GCE2 is a powerful method for self-consistently modelling the full dynamics of a range of spectral scales, coupled with the statistics of the quasilinear dynamics of the smaller spectral scales.} \NEW{An investigation of the GCE2 approximation at high resolution should be illuminating.}

\bigskip
Declaration of Interests. The authors report no conflict of interest.

\begin{acknowledgments}
We acknowledge support of funding from the European Union Horizon 2020 research and innovation programme (grant agreement no. D5S-DLV-786780). JBM and SMT are supported in part by a grant from the Simons Foundation (Grant No. 662962, GF).
\end{acknowledgments}

\bibliographystyle{jfm}
\bibliography{gce2}
\end{document}